%% file: Main.tex
\documentclass{aastex62}
\usepackage{amsmath}	
\usepackage{amssymb}	
\usepackage{natbib} 

\graphicspath{{./}{figures/}}

\received{May 8, 2020}
\accepted{June 14, 2020}

\submitjournal{ApJL}

\shorttitle{Sub-millimetre dimming of Betelgeuse}
\shortauthors{Dharmawardena et al.}

\begin{document}

\title{Betelgeuse fainter in the sub-millimetre too: \\
an analysis of JCMT and APEX monitoring during the recent optical minimum}

\correspondingauthor{Thavisha E. Dharmawardena}
\email{dharmawardena@mpia.de}

\author{Thavisha E. Dharmawardena}
\affiliation{Max-Planck-Institute for Astronomy, K\"onigstuhl 17, 69117 Heidelberg, Germany.}

\author{Steve Mairs}
\affiliation{East Asian Observatory,
660 N. A`oh\={o}k\={u} Place, Hilo, HI 96720, USA}

\author{Peter Scicluna}
\affiliation{European Southern Observatory, Alonso de Cordova 3107, Santiago RM, Chile}

\author{Graham Bell}
\affiliation{East Asian Observatory,
660 N. A`oh\={o}k\={u} Place, Hilo, HI 96720, USA}

\author{Iain McDonald}
\affiliation{Jodrell Bank Centre for Astrophysics, Alan Turing Building, University of Manchester, M13 9PL, UK}
\affiliation{Open University, Walton Hall, Milton Keynes, MK7 6AA, UK}

\author{Karl Menten}
\affiliation{Max-Planck-Institute for Radio Astronomy, Auf dem H{\" u}gel 69, 53121 Bonn, Germany}

\author{Axel Weiss}
\affiliation{Max-Planck-Institute for Radio Astronomy, Auf dem H{\" u}gel 69, 53121 Bonn, Germany}

\author{Albert Zijlstra}
\affiliation{Jodrell Bank Centre for Astrophysics, Alan Turing Building, University of Manchester, M13 9PL, UK}

\begin{abstract}
Betelgeuse, the nearest Red Supergiant star to us underwent an unusually deep minimum at optical wavelengths during its most recent pulsation cycle. We present submillimetre observations taken by the James Clerk Maxwell Telescope and Atacama Pathfinder Experiment over a time span of 13 years including the optical dimming. We find that Betelgeuse has also dimmed by $\sim$ 20\% at these longer wavelengths during this optical minimum. Using radiative-transfer models, we show that this is likely due to changes in the photosphere (luminosity) of the star as opposed to the surrounding dust as was previously suggested in the literature.
\end{abstract}


\section{Introduction} 
\label{sec:intro}

Massive stars (M $\geq$ 8\,M$_{\odot}$) are the main drivers of the chemical evolution in the Universe \citep{KarakasLattanzio2014PASA...31...30K}. These stars explode as supernovae, releasing nuclear-processes material into the interstellar medium, chemically enriching the surrounding environment. Between the main sequence and their explosive demise, massive stars undergo enhanced mass loss; since the mass-loss rate exceeds the nuclear-burning rate, mass lost during these phases becomes the determining factor in their evolution from this point onwards, determining what kind of supernova they will become \citep[e.g.][]{2012A&A...538L...8G,2013A&A...550L...7G}. For stars with masses $<$\,30\,M$_\odot$, the majority of this mass loss occurs as a red supergiant \citep[RSG; ][]{2013EAS....60..307V}.

The mechanisms driving RSG mass loss remain debatable, but pulsations are believed to play a role in at least some cases \citep[e.g.][]{2005A&A...438..273V,2009ApJ...701.1464H}. High-amplitude, long-period pulsations carry stellar surface material into the interstellar medium with the aid of strong stellar winds \citep{McDonaldTrabucchi2019MNRAS.484.4678M}. These semi-regular pulsations have periods spanning $\sim200 - 1000$ days are an inherent feature in evolved stars visible across a wide wavelength range \citep{Hofner+Olofsson2018}. 

\object{Betelgeuse}, the closest \citep[$152\pm20$ pc;][]{vanLeeuwen2007_Hipparcos} and best studied RSG, recently experienced an unusually deep minimum in its visual light curve \citep[e.g][]{ATEL1_2019ATel13341....1G, ATEL5_2020ATel13512....1G}, which captured both professional and public interest. Four main scenarios were put forward for this dimming; i. a confluence of the short ($\sim 400$ days) and long ($\sim 5$ years) periods; ii. changes in known hot and cold spots on the stellar surface; iii. a large ejection of newly formed dust along the line of sight; and more exciting iv. photospheric structural changes indicating an imminent supernova, which is now ruled out by the subsequent return to its original brightness.   

Some of these hypotheses have been examined in recent literature. 
Optical imaging of the star with VLT/SPHERE showed a distinct change in the apparent shape of the photosphere between January and December 2019, with the southern half of the star substantially dimmer in the December observations \footnote{see the recent ESO Photo release \url{https://www.eso.org/public/news/eso2003/}, data from M. Montarges}.
\citet{Levesque2020} compared spectra from 2004 and 2020, and interpreted the moderate change in TiO bands as evidence for a lack of change in effective temperature based on comparison to 1D static models.
Based on these considerations, the formation of a new dust cloud along the line of sight has emerged as the favoured hypothesis.

In this letter we present sub-millimetre (sub-mm) observations at $450~\micron$ and $850~\micron$ obtained by the Sub-mm Common User Bolometer Array 2 \citep[SCUBA-2,][]{Holland2013} instrument on the James Clerk Maxwell Telescope (JCMT). At these wavelengths, we avoid the effects of extinction along the line of sight, providing an unbiased probe of the emission of the star and its environs. Newly formed dust should be visible as increased emission, while the bright photosphere allows us to see any reduction in the overall luminosity of the star.

\section{Observations and Data Reduction}
\label{sec: Obs_DataRed}

JCMT/SCUBA-2 continuum observations were obtained as part of a director's discretionary time program and an urgent queue program (project IDs M19BD002 and S20AP001, respectively; PI: Mairs) at $450~\micron$ and $850~\micron$. Observations were carried out on 2020 Jan 23, Feb 16 and Mar 03 UT. These observations were approximately 11 minutes with $\tau_{225\mathrm{GHz}} \lesssim 0.05$, reaching noise levels of 74--103 and 16--27 mJy/beam for $450~\micron$ and $850~\micron$ respectively. Five archival JCMT/SCUBA-2 observations obtained in 2012 and 2013 were also included for comparison, with integration times of $\sim 6$ minutes with $\tau_{225\mathrm{GHz}} \lesssim 0.08$, reaching a noise level of 17--67 mJy/beam for $850~\micron$. 
The SCUBA-2 beam sizes at 850 and 450\,$\mu$m are $\sim 13''$ and $\sim 7.9''$ respectively.

All observations were reduced using {\sc makemap} \citep{Chapin2013} with an edited version of the {\sc dimmconfig\_bright\_compact} pipeline available in the JCMT reduction software {\sc starlink} version 2018A \citep{Currie2014}. The modified parameters in the  pipeline are as follows: {\sc numiter}=$-$200; {\sc flt.filt\_edge\_largescale} = 200; {\sc flt.zero\_circle} = $2\arcmin$; {\sc ast.zero\_circle} = $2\arcmin$ and {\sc maptol} = 0.005. In addition to these parameters, each observation was convolved with a $7\arcsec$ Gaussian to ensure that all the flux was recovered \citep{Dharmawardena2018, Dharmawardena2019B}. Flux-calibration factors were applied following the East Asian Observatory (EAO) guidelines\footnote{\url{https://www.eaobservatory.org/jcmt/instrumentation/continuum/scuba-2/calibration/}}.
The uncertainty on the absolute calibration of SCUBA-2 data is approximately 8\% \citep{Dempsey2013}.

The CO J=3--2 transition line at 345.7 GHz contaminates SCUBA-2 $850~\micron$ observations \citep{Drabek2012_SCUBA2_COcontam, Coude2016COSub, Dharmawardena2019A}. To remove this contamination from our observations, we use an archival CO 3--2 staring observation obtained from the JCMT/Heterodyne Array Receiver Program (HARP) instrument on 2013 December 29. The measured CO 3--2 contribution to the SCUBA-2 point source flux is 0.033\,Jy, for the half-power bandwidth of SCUBA-2 of 35\,GHz. 
The upper energy level of the CO 3--2 is $\approx 33$ K above the ground state, meaning that it samples the bulk gas in the envelope and hence we do not expect it to be strongly affected by variations in the temperature or luminosity of the star. 

The full list of observations is presented in Table~\ref{tab:flux}. Further, the SCUBA-2 data are complemented with archival $870~\micron$ data obtained with the Large APEX Bolometer Camera \citep{Siringo_Laboca_2009A&A...497..945S} which is operated on the Atacama Pathfinder Experiment submillimetre telescope (APEX), also shown in the same table. 
The  APEX/LABOCA 870\,$\mu$m observations of Betelgeuse were conducted between 2007 December and 2017 September.
All observations were carried out using standard raster-spiral observations \citep{Siringo_Laboca_2009A&A...497..945S} under good weather conditions (PWV=0.6--1.4\,mm).
Integration times were between 180 and 750 seconds. Flux calibration was achieved through
observations of Mars, Uranus, Neptune and secondary calibrators and is typically accurate 
within 8.5\% rms. The atmospheric attenuation was determined via skydips every 2hr and from independent data from the APEX radiometer, which measures the line-of-sight
water-vapour column every minute. The data were reduced and imaged using the BoA reduction
package as detailed in \citet{Weiss_Boa_2009ApJ...707.1201W}. LABOCA's central frequency and
beam size are 345\,GHz and 19.2$''$. Fluxes were derived from the peak flux densities on
the maps after smoothing the data with a Gaussian with a FWHM of 12$''$. Resulting noise 
levels are between 10 and 30\,mJy/beam.

\input{fluxtable}

\newpage

\section{Analysis and Discussion}
\label{sec: Analysis_Disc}

\subsection{Sub-mm flux variation}
We derive the SCUBA-2 point source photometry for each observation by integrating under the point-spread function (PSF) scaled to the peak flux at sub-pixel precision. The PSF is generated by convolving the standard SCUBA-2 PSF consisting of two Gaussian components \citep{Dempsey2013} and the $7''$ smoothing Gaussian. This approach is more effective in order to measure the compact-flux component, avoiding contamination by any negative bowling effects, and by the extended circumstellar-envelope component, which methods such as aperture photometry may cause. The pre-optical dimming (2012 -- 2013) JCMT/SCUBA-2 $850~\micron$ fluxes are consistent with those derived by \citet{OGorman2017_ALMABetel} using the ALMA main array at the same wavelength. 

\begin{figure}
    \centering
    \includegraphics[width=0.7\textwidth]{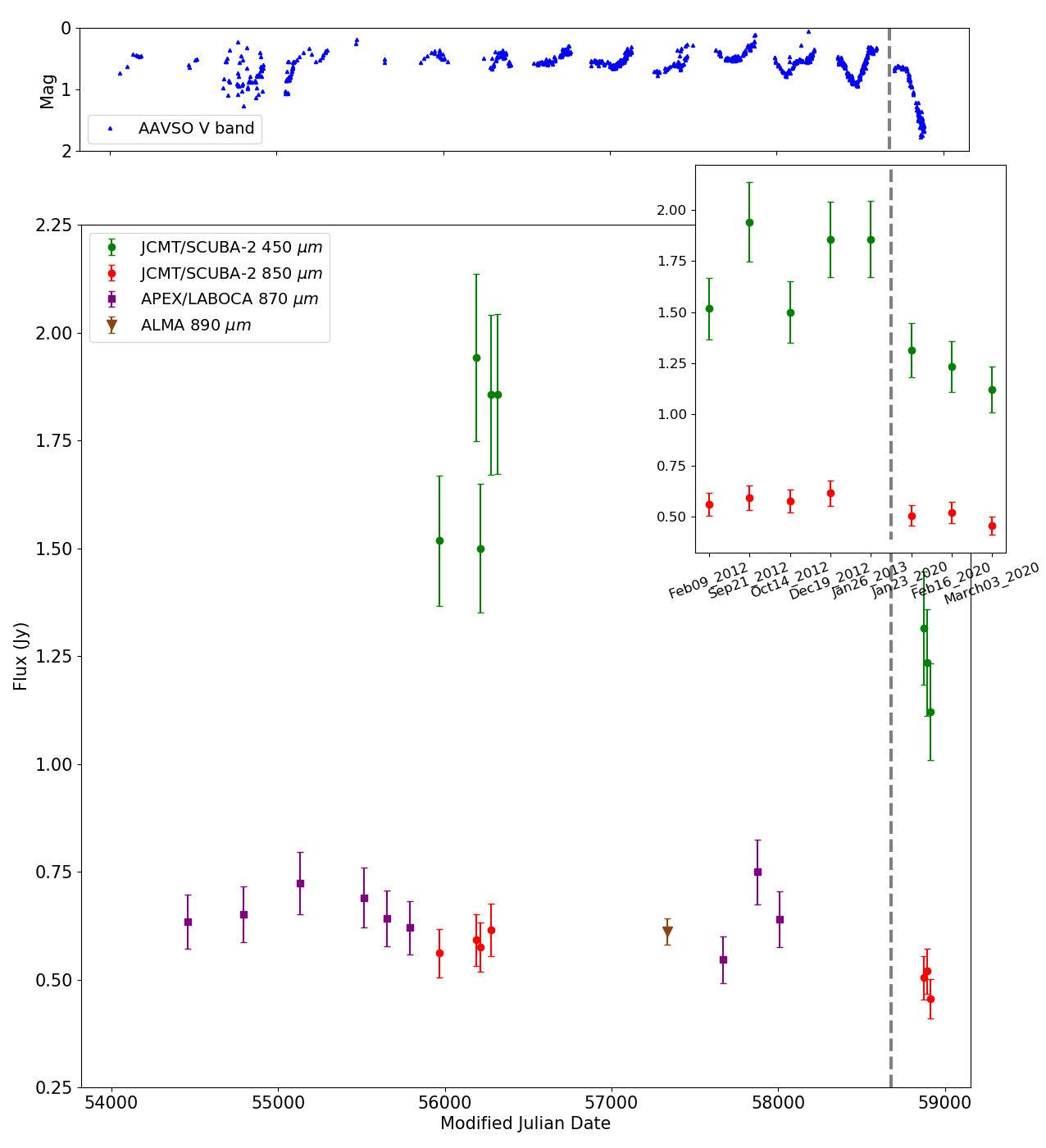}
    \caption{Lightcurves of Betelgeuse for the last fifteen years: The top panel shows the AAVSO optical (V band) light curve; The bottom panel shows the JCMT/SCUBA-2 $450~\micron$ (green points) and $850~\micron$ (red points) light curve constructed using both archival and newly obtained observations. Here we also show the the APEX/LABOCA $870~\micron$ light curve (purple squares) obtained from archival data. While not used in the fitting we also show the single ALMA $890~\micron$ data point (brown triangle) from \citet{OGorman2017_ALMABetel}, to illustrate the consistency. The grey dashed line indicates the beginning of the dimming of the recent pulsation cycle. The inset panel shows a zoomed in version of the JCMT/SCUBA-2 light curve with the corresponding UT date for each observation. {\bf All error bars indicate 1-$\sigma$ uncertainties including the uncertainty on the absolute calibration.}}
    \label{fig:lightcurve}
\end{figure}

The JCMT/SCUBA-2 light curve, along with the archival APEX/LABOCA data, is shown in Fig.\ref{fig:lightcurve}. For comparison, the AAVSO\footnote{\url{https://www.aavso.org}} optical (V band) light curve for the same date range is also shown here. 

\subsection{Modelling the lightcurve}\label{sec:model}
To determine whether there is any evidence of variation in the sub-mm fluxes of Betelgeuse, we perform Bayesian inference by forward-modelling the fluxes with three different models and evaluate which one best-reproduces the fluxes simultaneously for SCUBA-2 850, SCUBA-2 450 and LABOCA. {\bf Because the stellar signal is strong, we assume that the uncertainties of the fluxes are dominated by the systematic, uncorrelated scatter in the determination of the absolute calibration \citep{Dempsey2013}. The models we consider are:}

\begin{description}
    \item [C1] constant flux. This model has three free parameters: $c$, the SCUBA-2 850\,$\mu$m flux at 850\,$\mu$m; $\alpha_{450}$, the ratio between the 450 and 850 fluxes; and $\alpha_{870}$, the ratio between the SCUBA-2 850 and LABOCA 870\,$\mu$m fluxes. These $\alpha$ parameters allow us to straightforwardly handle the conversion from the calibration of one instrument to another, which would otherwise inject additional uncertainty on the value of $c$.
    \item [C2] the behaviour is broken into two distinct epochs each with constant flux. Unlike model C1, only the $\alpha$ parameters are assumed to be fixed at all times. There are then three other parameters, $t_{\rm break}$, the date at which the behaviour changes, and $c$ and $c^{\prime}$, the (otherwise constant) SCUBA-2 850\,$\mu$m flux before and after $t_{\rm break}$, respectively.
    \item [L] the flux varies linearly with time, with all bands following the slope. This is similar to model C1, except that an additional parameter $m$ is added so that the fluxes are proportional to $m t + c$. This effectively assumes that the fractional rate of change at each band is the same.
\end{description}
In the absence of further constraints, the priors on all parameters are assumed to be flat, drawn from uniform random distributions that cover the required ranges, listed in Table~\ref{tab:model}. 

We now have a model-selection problem including multiple comparisons. 
To identify the most appropriate model of the three, we must compute Bayes' factors ($K$) for each comparison.
{\bf The Bayes’ factor provides a way to select between families of models by considering the total evidence (or integrated likelihood) over the whole prior volume and comparing the ratio of evidences for different families of models. 
This} requires that we have a reliable way of estimating the \textit{evidence}, $Z$ for each model \citep[for further details on Bayes' factors, see e.g.][]{Goodman1999_BF1, Goodman1999_BF2, MOREY2016_BF3}. 
To do this, we use the Python nested-sampling package {\sc dynesty}, which supports dynamic nested sampling, allowing optimisation for either evidence estimation or posterior exploration \citep{SkillingNS2004AIPC..735..395S,skillingNS2006,HigsonDNS2019S&C....29..891H,dynestySpeagle2020MNRAS.493.3132S}.
For each model, we conduct two runs, one optimised to estimate the evidence for the model selection, and the other optimised to evaluate the posterior and hence provide parameter estimates.
All runs used 1500 initial live points and {\sc dlogz\_init} = 0.01.
The estimated evidence and parameter values for each model are given in Table~\ref{tab:model}, along with their respective $\mathbf{1\sigma}$ uncertainties.

\input{modeltable}

Using the evidence computed using {\sc dynesty}, we now compute Bayes' factors for each pair of models.
Unlike frequentist approaches, Bayes' factors automatically penalise models with more free parameters but do not privilege any model by considering it the null hypothesis.
By assuming that all models are \textit{a priori} equally likely, the Bayes factor collapses to the ratio of the evidences --- the impact of this assumption is difficult to quantify, but this is a necessary starting point.
The frequentist equivalent to the process would be to separately compare model C2 and L to C1, considering model C1 as the null hypothesis. 
However, this would require an additional penalty as we are performing multiple comparisons, while the use of Bayes factors avoids this issue.

The ratios of the evidence support model C2 over the other two models by a significant margin.
For the comparison between C2 and C1, the Bayes' factor $K = 710$, while the $K$ comparison between C2 and L is $\sim 6\times10^6$.
This represents \emph{decisive} evidence in favour of model C2 \citep[using the scale of e.g.][]{KassRaffertyBayesFactors}, suggesting not only that the sub-mm flux is variable but, crucially, that it is \emph{lower} during the recent minimum than it was before.
Given that the sub-mm emission is dominated by the star, this implies that the luminosity has decreased by nearly 20\% {\bf (a $3.6\sigma$ change)} and argues against models in which the recent optical dimming is caused by the formation of a new cloud of dust along the line of sight to the star, as increases in the dust mass \emph{cannot} decrease the sub-mm flux as the dust emission at this wavelength is optically thin\footnote{This is only true if the overall temperature structure does not change. In exceptional cases, it is possible to envisage the optical depth increasing sufficiently to lower the dust temperature such that the emission is reduced, but the radiative-transfer modelling shown in this section makes this a moot point in the case of Betelgeuse.}.

To understand the role of dust versus the photosphere in the dimming across the optical--sub-mm wavelength range, we perform dust radiative-transfer modelling.
We do not aim to quantitatively fit the observations, merely to qualitatively illustrate the trends that could be expected under different assumptions.
Using the Monte Carlo radiative-transfer code {\sc hyperion} \citep{Robitaille2011_Hyperion}, we compute models of a spherically-symmetric wind with constant outflow speed of 14\,km\,s$^{-1}$ \citep[comparable to ][]{Loup1993A&AS...99..291L}.
We use dust that consists of compact spherical grains composed of 30\% alumina using optical constants from \citet{Begemann1997ApJ...476..199B} and 70\% Mg-Fe silicate for optical constants from \citet{Dorschner1995A&A...300..503D}, with grain sizes from 0.1--1~$\mu$m following an MRN size distribution \citep{MRN1977ApJ...217..425M}.
We use one million photons per iteration, which is sufficient as the shell is optically thin and spherically symmetric, with grid cells distributed logarithmically in radius. 
Using a dust mass-loss rate of $3\times 10^{-9}$~M$_{\odot}$\,yr$^{-1}$ \citep[e.g.][]{Verhoelst2009} we achieve an acceptable fit to the optical--mid-infrared photometry.
We then compute a second model, in which an additional shell of dust has been added between 2 and 4 R$_{\ast}$, close to the observed dust-condensation radius \citep{Haubois2019A&A...628A.101H}, which adds an additional 1 magnitude of optical extinction.

\begin{figure}
    \centering
    \includegraphics[width=0.7\textwidth]{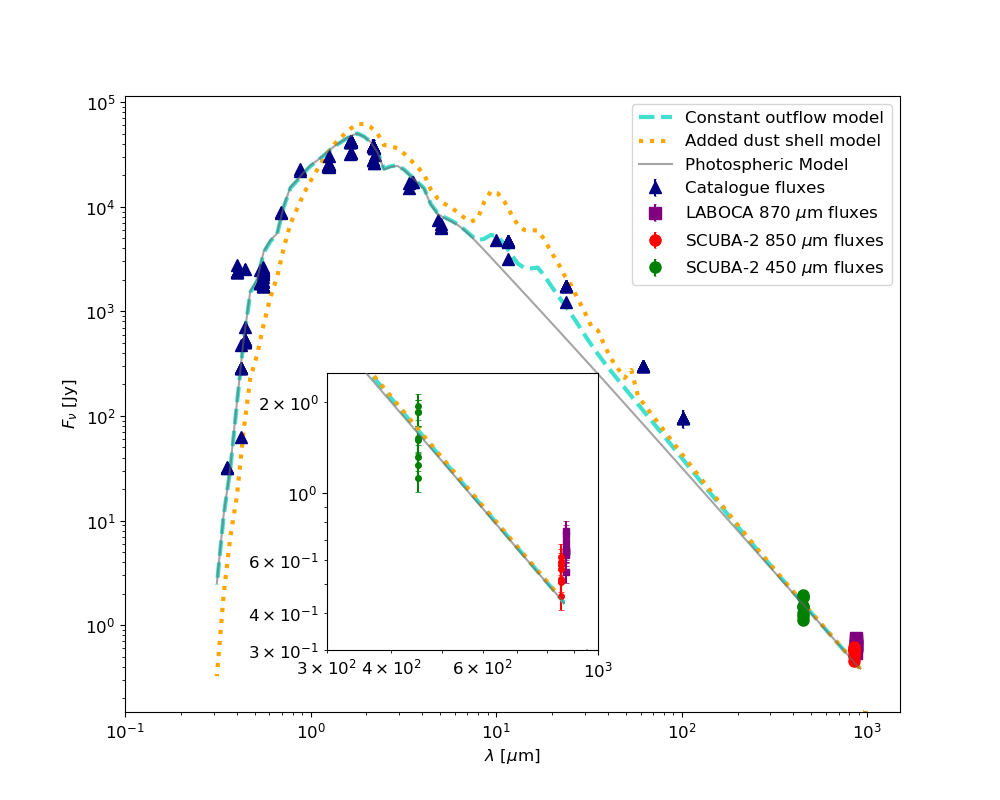}
    \caption{Radiative-transfer models (lines) compared with observations of Betelgeuse. Blue triangles show fluxes, taken from Vizier (listed in the online appendix, Table~\ref{tab:vizfluxes}), and the green, red and purple points indicate the data from this work in the same colour scheme as Fig.~\ref{fig:lightcurve}. The blue dashed line represents a qualitative fit from {\sc Hyperion} to the 0.3--30 $\mu$m photometry, and the orange dotted line a model that is identical other than the addition of an extra shell of dust from 2--4R$_\ast$ that produces 1 magnitude of V-band extinction and the grey solid line shows the underlying photospheric model at the same spectral resolution as the radiative-transfer models. The inset panel highlights the sub-mm fluxes from this work and the model flux in the same region. }
    \label{fig:RTmodels}
\end{figure}

Comparing these two models (see Fig.~\ref{fig:RTmodels}), we see two distinctive features.
Firstly, while the addition of new dust substantially changes the optical and infrared emission, it makes very little difference at wavelengths longer than 100~$\mu$m.
Secondly, Betelgeuse's emission at sub-mm wavelengths is entirely dominated by the stellar photosphere, with small contributions from circumstellar dust and the radio \textbf{pseudo-photosphere (i.e. optically-thick excess radio emission whose size changes with wavelength,} not included in our models) \citep[e.g.][]{Richards2013MNRAS.432L..61R, OGorman2017_ALMABetel}.
The radio pseudo-photosphere, which dominates the spectrum at cm wavelengths and beyond \citep{Ogorman2015}, clearly makes only a minor contribution at 850\,$\mu$m and a negligible one at 450\,$\mu$m.
Therefore, while we are unable to probe changes in dust mass at the level that would produce the optical dimming, it is clear that only changes in the photosphere can reproduce the sub-mm dimming -- this would otherwise require that the dust emission or the radio pseudo-photosphere had completely disappeared.
Using the same logic, we can exclude changes in the sub-mm line emission as a cause of the variations.
Similar to CW Leo \citep{Dharmawardena2019A}, the CO contribution (the brightest line in these spectral regions) to the measured continuum is at the 5--10\% level; short of all the molecular line emission disappearing, the continuum could not change by this amount.

\begin{figure}
    \centering
    \includegraphics[width=\textwidth, clip=true, trim=2cm 0cm 3cm 1cm]{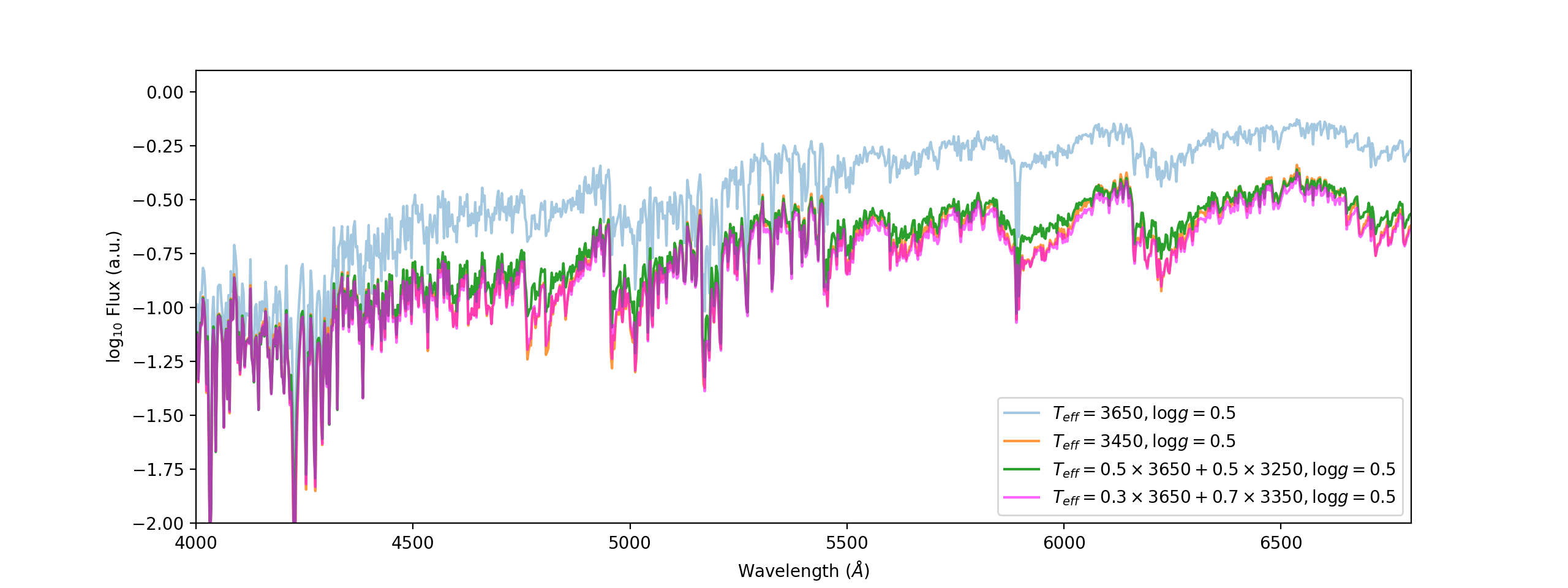}
    \caption{\bf Comparison of optical (4000--7000~\AA) model spectra (PHOENIX spectra \citep{PHOENIX_2013A&A...553A...6H} interpolated using the {\sc starfish} package \citep{starfishpaper_2015ApJ...812..128C,starfishcode_ian_czekala_2018_2221006}) showing the influence of temperature changes and mixing of different temperature components, normalised so that the spectra are equal to one at the peak of the SED ($\sim 1.5$\,\micron) with the same wavelength sampling as in \citet{Levesque2020}, although not convolved with the line response function. All three scenarios outlined in the text have rather similar spectra, and are qualititatively similar to Figure 1 in \citet{Levesque2020}.}
    \label{fig:spec_comp}
\end{figure}

Under the na\"ive assumption that the change in sub-mm flux corresponds to a change in temperature at constant radius, and that the original temperature was $T_{\rm eff} = 3650$\,K \citep[the 2004 temperature from][]{Levesque2020}, this would suggest a present value of 3450\,K.
Alternatively, this could be reconciled with the sub-mm data if starspots with $T_{\rm eff} = 3250$\,K cover 50\% of the visible surface, or if spots with $T_{\rm eff} = 3350$\,K cover 70\% of the visible surface. 
{\bf Indeed, as can be seen in Fig.~\ref{fig:spec_comp}, all such scenarios would also be accompanied by a $\sim 0.9$\,mag dimming in the V band, and are qualitatively similar to the 2020 spectrum of \citet{Levesque2020}, including the convergence of the spectra at the blue end with the warmer/2004 spectrum.}
However, as a dynamic, non-equilibrium system, changes in temperature and the resulting molecular bands in pulsating stars may take some time to settle after a significant perturbation \citep[e.g.][]{McDonaldvanLoon2007, Lebzelter2010A&A...517A...6L, Lebzelter2014A&A...567A.143L}.
Hence, one-dimensional models have difficulty capturing the observed behaviour and detailed 3D modelling is required.
This might explain why the expected changes in TiO bands were not visible in the spectra of \citet{Levesque2020}, as it is typical to compare to static, one-dimensional models.
If the change were instead due to a change in radius at constant temperature, the change in radius would be small ($\sim10\%$), since $\Delta R\propto\sqrt{\Delta L}$.

\section{Conclusions}\label{sec:conc}
We present new and archival sub-mm observations from JCMT/SCUBA-2 and APEX/LABOCA of the red supergiant Betelgeuse during its recent deep optical minimum. 
Modelling the sub-mm lightcurve shows that it is inconsistent with the sub-mm flux remaining constant during the recent dimming.
The models suggest that the sub-mm flux has declined by 20\% compared to pre-dimming values. 
Radiative-transfer modelling shows that this cannot be caused by changes in the dust envelope, and therefore must reflect a change in the photosphere of the star.
This 20\% reduction in luminosity could take the form of a change in radius or temperature.
The required change in radius would be small ($\sim10\%)$, while a change in temperature could be explained either through a $\sim$200\,K global change, or through the presence of spots $\sim$ 400\,K cooler covering $\sim50\%$ of the visible surface.

\section*{Acknowledgements}

We would like to thank Amelia Bayo for sharing LABOCA obserations and Sarah Graves for help determining the CO contribution to SCUBA-2.

We acknowledge with thanks the variable star observations from the AAVSO International Database contributed by observers worldwide and used in this research.

This project is partially funded by the Sonderforschungsbereich SFB881 “The Milky Way System” of the German Research Foundation (DFG).

The James Clerk Maxwell Telescope is operated by the East Asian Observatory on behalf of The National Astronomical Observatory of Japan; Academia Sinica Institute of Astronomy and Astrophysics; the Korea Astronomy and Space Science Institute; the Operation, Maintenance and Upgrading Fund for Astronomical Telescopes and Facility Instruments, budgeted from the Ministry of Finance (MOF) of China and administrated by the Chinese Academy of Sciences (CAS), as well as the National Key R\&D Program of China (No. 2017YFA0402700). Additional funding support is provided by the Science and Technology Facilities Council of the United Kingdom and participating universities in the United Kingdom and Canada. Program IDs: M19BD002, S20AP001, M10AEC30, M11BEC30, M12AEC05, M12BEC05 and M13BN01. 

This publication is based on data acquired with the Atacama Pathfinder Experiment (APEX). APEX is a collaboration between the Max-Planck-Institut fur Radio astronomie, the European Southern Observatory, and the Onsala Space Observatory. Observation project IDs M-79.F-0147-2007, M-081.F-1039-2008, M-083.F-0057-2009, M-085.F-0066-2010,
C-098.F-9705A-2016, C-099.F-9720A-2017 and M-099.F-9529C-2017

\bibliographystyle{aasjournal}
\bibliography{Betelgeuse_Bib}

\appendix
\section{Literature fluxes used for radiative-transfer modelling}
\input{Betelgeuse_photometry_for_appendix_table}
\end{document}

%% file: fluxtable.tex
\begin{table}[]
    \centering
    \caption{Table of fluxes}\label{tab:flux}
    \begin{tabular}{llccc}
        \hline MJD & UT Date & SCUBA-2 850 flux & SCUBA-2 450 flux & LABOCA 870 flux \\ 
         &  & (Jy) & (Jy) & (Jy) \\\hline\hline
        54458 & 2007--12--24 & -- & -- & 0.634 \\
        54791 & 2008--11-21 & -- & -- & 0.651 \\
        55132 & 2009--10--28 & -- & --  & 0.724 \\
        55517 & 2010--11-17 & --  & --  & 0.690 \\
        55654 & 2011--04--03 & --  & --  & 0.642 \\
        55792 & 2011-08--19 & --  & --  & 0.620 \\
        55966 & 2012--02--09 & 0.573 & 1.52 & -- \\
        56191 & 2012--09--21 & 0.603 & 1.94 & -- \\
        56214 & 2012--10--14 & 0.587 & 1.50 & --  \\
        56280 & 2012--12--19 & 0.627 & 1.86 & -- \\
        56318 & 2013--01--26 & 0.483 & 1.86 & -- \\
        57668 & 2016--10--07 & --  & --  & 0.546 \\
        57874 & 2017--05-01 & --  & --  & 0.750 \\
        58007 & 2017--09--11 & --  & --  & 0.640 \\
        \hline
        58871 & 2020--01--23 & 0.517 & 1.32 & -- \\
        58895 & 2020--02--16 & 0.531 & 1.23 & -- \\
        58911 & 2020--03--03 & 0.467 & 1.12 & -- \\
        \hline
    \end{tabular}
\end{table}

%% file: modeltable.tex
\begin{table}[]
    \centering
    \caption{Results of fitting using nested sampling}
    \label{tab:model}
    \begin{tabular}{lccccccc}
        \hline Model & $\ln{Z}$ & $\alpha_{870}$ & $\alpha_{450}$ & $c$ & $c^{\prime}$ & $m$ & $t_{\rm break}$ \\
         & & & & [Jy] & [Jy]& & MJD \\
        \hline \hline
         C1 &$-36.11\pm0.06$&$1.23\pm0.05$&$2.73\pm0.13$&$0.525\pm0.015$&--&--& --\\ 
         C2 &$-29.24\pm0.07$&$1.15\pm0.06$&$2.81\pm0.13$&$0.567\pm0.018$&$0.466\pm0.021$&--&$58320^{+380}_{-900}$ \\
         L &$-44.84\pm0.11$&$1.15\pm0.05$&$2.79\pm0.13$&$2.22\pm0.41$&--&$(-2.96\pm0.72)\times10^{-5}$&-- \\ \hline
         Priors & -- & 0 -- 2 & 0 -- 10 & C1, C2: 0 -- 1 & 0 -- 1 & $-$3 -- 3 & 56000 -- 59000 \\
          & & & & L: 0 -- 100 & & & \\\hline
    \end{tabular}
\end{table}

%% file: Betelgeuse_photometry_for_appendix_table.tex
\begin{table}
\centering\caption{Fluxes used in Fig.~\ref{fig:RTmodels}}\label{tab:vizfluxes}
\footnotesize
  \begin{tabular}{lcccl}
    \hline Filter & Freq & Flux & Uncertainty & Reference Vizier table \\
    & $\mathrm{GHz}$ & $\mathrm{Jy}$ & $\mathrm{Jy}$ &  \\\hline\hline
    Johnson:U & 849030 & 32.6 &  & II/7A/catalog \\
    Johnson:U & 849030 & 32 &  & II/7A/catalog \\
    HIP:Hp & 745750 & 2.37$\times 10^3$ & 50 & I/239/hip\_main \\
    HIP:Hp & 745750 & 2.77$\times 10^3$ &  & V/137D/XHIP \\
    HIP:BT & 713280 & 286 & 2 & I/239/hip\_main \\
    HIP:BT & 713280 & 62.5 &  & I/275/ac2002 \\
    Johnson:B & 674900 & 2.55$\times 10^3$ & 2.8$\times 10^2$ & IV/38/tic \\
    Johnson:B & 674900 & 517 &  & II/7A/catalog \\
    Johnson:B & 674900 & 531 &  & II/7A/catalog \\
    Johnson:B & 674900 & 507 & 9 & I/305/out \\
    Johnson:B & 674900 & 707 &  & V/137D/XHIP \\
    Johnson:B & 674900 & 507 & 8 & V/145/sky2kv5 \\
    Johnson:B & 674900 & 527 &  & B/pastel/pastel \\
    HIP:VT & 563630 & 1.85$\times 10^3$ & 30 & I/239/hip\_main \\
    Johnson:V & 541430 & 1.93$\times 10^3$ &  & II/122B/merged \\
    Johnson:V & 541430 & 2.47$\times 10^3$ &  & II/7A/catalog \\
    Johnson:V & 541430 & 2.15$\times 10^3$ &  & I/239/tyc\_main \\
    Johnson:V & 541430 & 2.52$\times 10^3$ &  & II/7A/catalog \\
    Johnson:V & 541430 & 1.91$\times 10^3$ &  & II/21A/catalog \\
    Johnson:V & 541430 & 1.74$\times 10^3$ &  & II/53/catalog \\
    Johnson:V & 541430 & 2.4$\times 10^3$ &  & J/other/NatAs/4.03/lit \\
    Johnson:V & 541430 & 2.52$\times 10^3$ &  & II/122B/merged \\
    Johnson:V & 541430 & 2.3$\times 10^3$ &  & III/124/stars \\
    Johnson:V & 541430 & 2.47$\times 10^3$ &  & B/pastel/pastel \\
    Johnson:V & 541430 & 2.65$\times 10^3$ & 60 & IV/38/tic \\
    Johnson:V & 541430 & 2.4$\times 10^3$ &  & I/239/hip\_main \\
    Johnson:R & 432100 & 8.65$\times 10^3$ &  & II/7A/catalog \\
    Johnson:R & 432100 & 8.89$\times 10^3$ &  & II/7A/catalog \\
    Johnson:I & 341450 & 2.28$\times 10^4$ &  & II/7A/catalog \\
    Johnson:I & 341450 & 2.22$\times 10^4$ &  & II/7A/catalog \\
    Johnson:J & 239830 & 2.53$\times 10^4$ & 2.4$\times 10^3$ & II/246/out \\
    Johnson:J & 239830 & 2.39$\times 10^4$ &  & II/7A/catalog \\
    Johnson:J & 239830 & 2.55$\times 10^4$ &  & II/7A/catalog \\
    DIRBE:1.25 & 237320 & 3.06$\times 10^4$ &  & J/ApJS/190/203/var \\
    Johnson:H & 183920 & 4.17$\times 10^4 $& 6.2$\times 10^3$ & II/246/out \\
    Johnson:H & 183920 & 3.23$\times 10^4$ &  & II/7A/catalog \\
    Johnson:K & 136890 & 3.68$\times 10^4$ & 6.3$\times 10^3$ & II/246/out \\
    Johnson:K & 136890 & 2.62$\times 10^4$ &  & II/7A/catalog \\
    Johnson:K & 136890 & 2.6$\times 10^4$ &  & II/7A/catalog \\
    DIRBE:2.2 & 134960 & 3.11$\times 10^4$ &  & J/ApJS/190/203/var \\
    Johnson:L & 88174 & 1.7$\times 10^4$ &  & II/7A/catalog \\
    Johnson:L & 88174 & 1.51$\times 10^4$ &  & II/7A/catalog \\
    DIRBE:3.5 & 85118 & 1.73$\times 10^4$ &  & J/ApJS/190/203/var \\
    DIRBE:4.9 & 61312 & 7.49$\times 10^3$ &  & J/ApJS/190/203/var \\
    Johnson:M & 59601 & 6.41$\times 10^3$ &  & II/7A/catalog \\
    Johnson:M & 59601 & 6.23$\times 10^3$ &  & II/7A/catalog \\
    Johnson:M & 59601 & 6.9$\times 10^3$ &  & II/7A/catalog \\
    :=10um & 29979 & 4.8$\times 10^3$ &  & II/53/catalog \\
    IRAS:12 & 25866 & 4.68$\times 10^3$ & 1.9$\times 10^2$ & II/125/main \\
    IRAS:25 & 12554 & 1.74$\times 10^3$ & 70 & II/125/main \\
    IRAS:60 & 4847.1 & 299 & 21 & II/125/main \\
    IRAS:100 & 2940.6 & 95.9 & 19 & II/125/main \\\hline
  \end{tabular}
\end{table}